\documentclass[epj]{webofc}
\usepackage[varg]{txfonts}   
\wocname{EPJ Web of Conferences}
\woctitle{CONF12}
\begin{document}
\selectlanguage{english}
\title{String point of view for heavy-light mesons\footnote{A talk given by T. Matsuki at XII Quark Confinement and the Hadron Spectrum.}}
%
%

\author{Yubing Dong\inst{1,2}\fnsep\thanks{\email{dongyb@ihep.ac.cn}} \and
        Qi-Fang L\"u\inst{1}\fnsep\thanks{\email{lvqifang@ihep.ac.cn}} \and
        Takayuki Matsuki\inst{3,4}\fnsep\thanks{\email{matsuki@tokyo-kasei.ac.jp}}
}

\institute{Institute of High Energy Physics, CAS, Beijing 100049, People's Republic of China 
\and
           Theoretical Physics Center for Science Facilities (TPCSF), CAS, People's Republic of China 
\and
           Tokyo Kasei University, 1-18-1 Kaga, Itabashi, Tokyo 173-8602, Japan
\and
           Theoretical Research Division, Nishina Center, RIKEN, Wako, Saitama 351-0198, Japan
}

\abstract{%
An approximate rotational symmetry of a heavy-light meson is viewed from a string picture. Using a simple string configuration, we derive a formula, $(M-m_c)^2=\pi\sigma L$, whose coefficient of the r.h.s. is just 1/2 of that of a light meson with two light quarks. A numerical plot is obtained for $D$ mesons of experimental data as well as several theoretical models, which shows good agreement with this formula.
}
\maketitle
\section{Observation of experimental spectra}
\label{intro}
In the former paper \cite{Matsuki:2016hzk}, we have pointed out that a careful observation of the experimental spectra of heavy-light mesons tells us that heavy-light mesons with the same angular momentum $L$ are almost degenerate and that mass differences within a heavy quark spin doublet and between doublets with the same $L$ are very small compared with a mass gap between different multiplets with different $L$, which is nearly equal to the value of the $\Lambda_{QCD}\sim 300$ MeV. 

In this report, we would like to give a different point of view to heavy-light mesons, i.e., hadronic open strings. With this picture, we can intuitively and physically see why masses of heavy-light mesons are proportional to $L$.

\section{Mass and Angular Momentum}\label{sec2}

We take Nambu's picture \cite{Nambu:1974zg} of a hadronic string, which consists only of gluons and both ends of which quarks are attached to in the case of mesons.
Afonin cited Nambu's paper to derive the Chew-Frautsch formula:
\begin{eqnarray}
  M^2 = 2\pi\sigma L. \label{eq:CF1}
\end{eqnarray}
Afonin's way of derivation is as follows.\cite{Afonin:2007jd} See also Refs. \cite{Afonin:2007aa,Afonin:2013hla} for other applications.
Assuming the simple string configuration, we obtain the mass $M$ and angular momentum $L$ given by the following equations,
\begin{eqnarray}
  M &=& 2\int_0^{\ell/2}\frac{\sigma dr}{\sqrt{1-v^2(r)}} = \frac{\pi\sigma \ell}{2}, \label{eq:ML1} \\
  L &=& 2\int_0^{\ell/2}\frac{\sigma rv(r)dr}{\sqrt{1-v^2(r)}}=\frac{\pi\sigma\ell^2}{8},
  \label{eq:ML2}
\end{eqnarray}
where $\sigma$ is a string tension, $\ell$ is a length of a string which connects two light quarks at the ends, and $v(r)=2r/\ell$ is the speed of the flux tube at the distance $r$ from the center of rotation.
Equations (\ref{eq:ML1}) and (\ref{eq:ML2}) are obtained by assuming the simplest configuration of a string which connects two quarks at both ends of a string with a speed of light $c$ and rotates around the center of the mass system. Combining these two equations, we arrive at Eq. (\ref{eq:CF1}).

\begin{figure}[htpb]
	\begin{center}
		\includegraphics[scale=0.5]{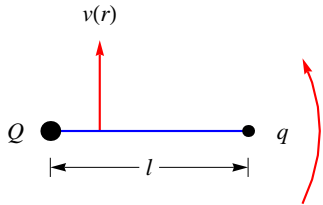}
		\caption{(color online). The schematic diagram for depicting the string connecting heavy and light quarks at the ends and rotating around the heavy quark. A heavy quark is fixed at one point.}\label{stringfig}
	\end{center}
\end{figure}

Let us apply the above idea to a heavy-light meson. In the heavy quark limit, one considers the situation that a heavy quark is fixed at one end and a light quark rotates around a heavy quark with a speed of light $c$ as described in Fig. \ref{stringfig}. Then one obtains the following relation:
\begin{eqnarray}
  M^2 = \pi\sigma L. \label{eq:CF}
\end{eqnarray}
The right hand side of this equation is just 1/2 of Eq. (\ref{eq:CF1}) and our numerical plot does not fit with this equation.
Because the Nambu's string consists only of gluons, we need to get rid of effects of quark masses. In our case, in the limit of heavy quark effective theory, we subtract only the $c$ quark mass from $M$. Hence the final expression we adopt is given by
\begin{eqnarray}
  \left(M-m_c\right)^2 = \pi\sigma L. \label{eq:CF2}
\end{eqnarray}

\begin{figure}[htpb]
	\begin{center}
		\includegraphics[scale=0.5]{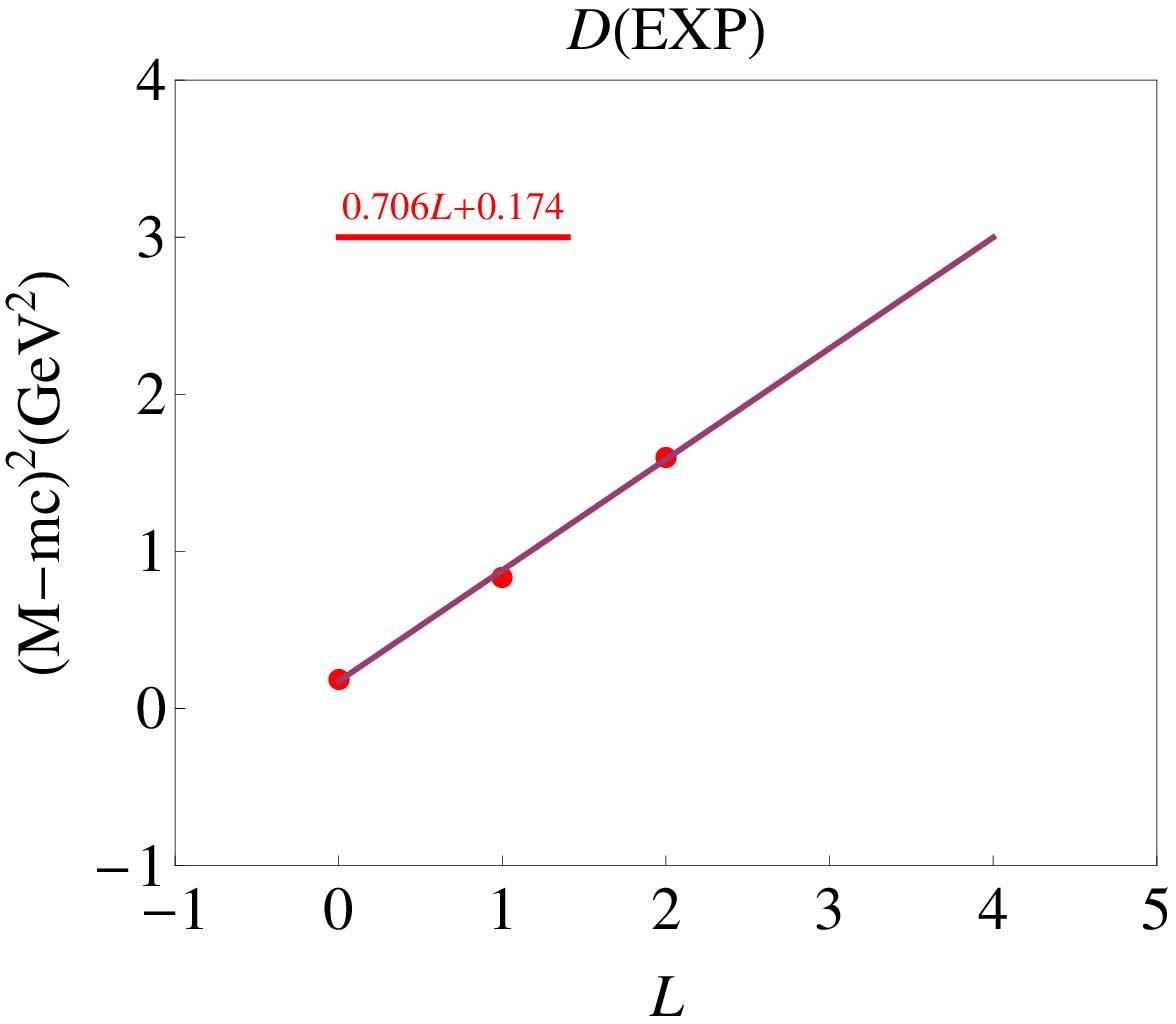} \;\;
		\includegraphics[scale=0.5]{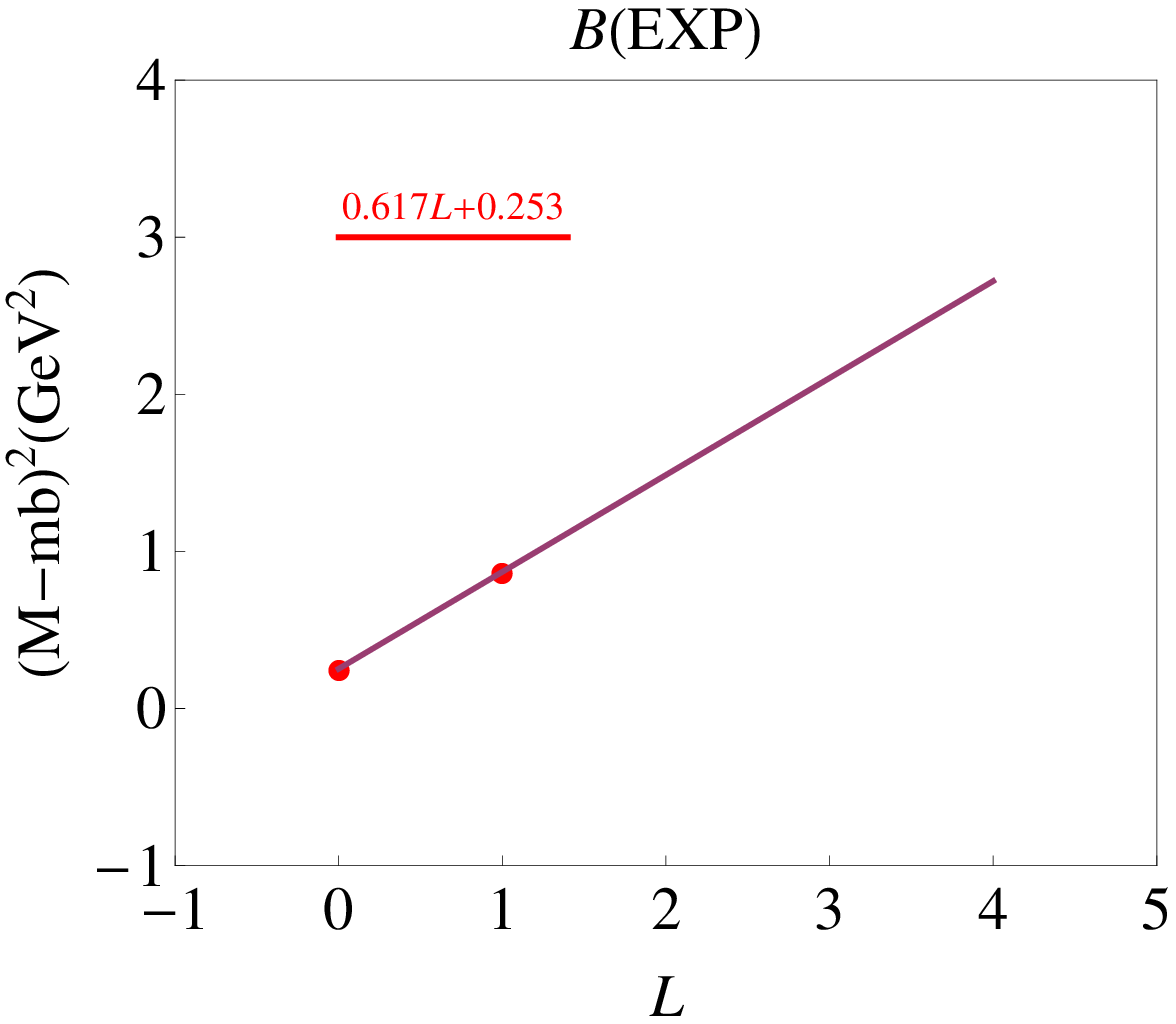}
		\caption{(color online). Plots of experimental data. $L$ versus $(M-m_{c,b})^2$. The best fit lines are given with equations.}\label{expfig}
	\end{center}
\end{figure}

\begin{figure}[htpb]
	\begin{center}
		\includegraphics[scale=0.5]{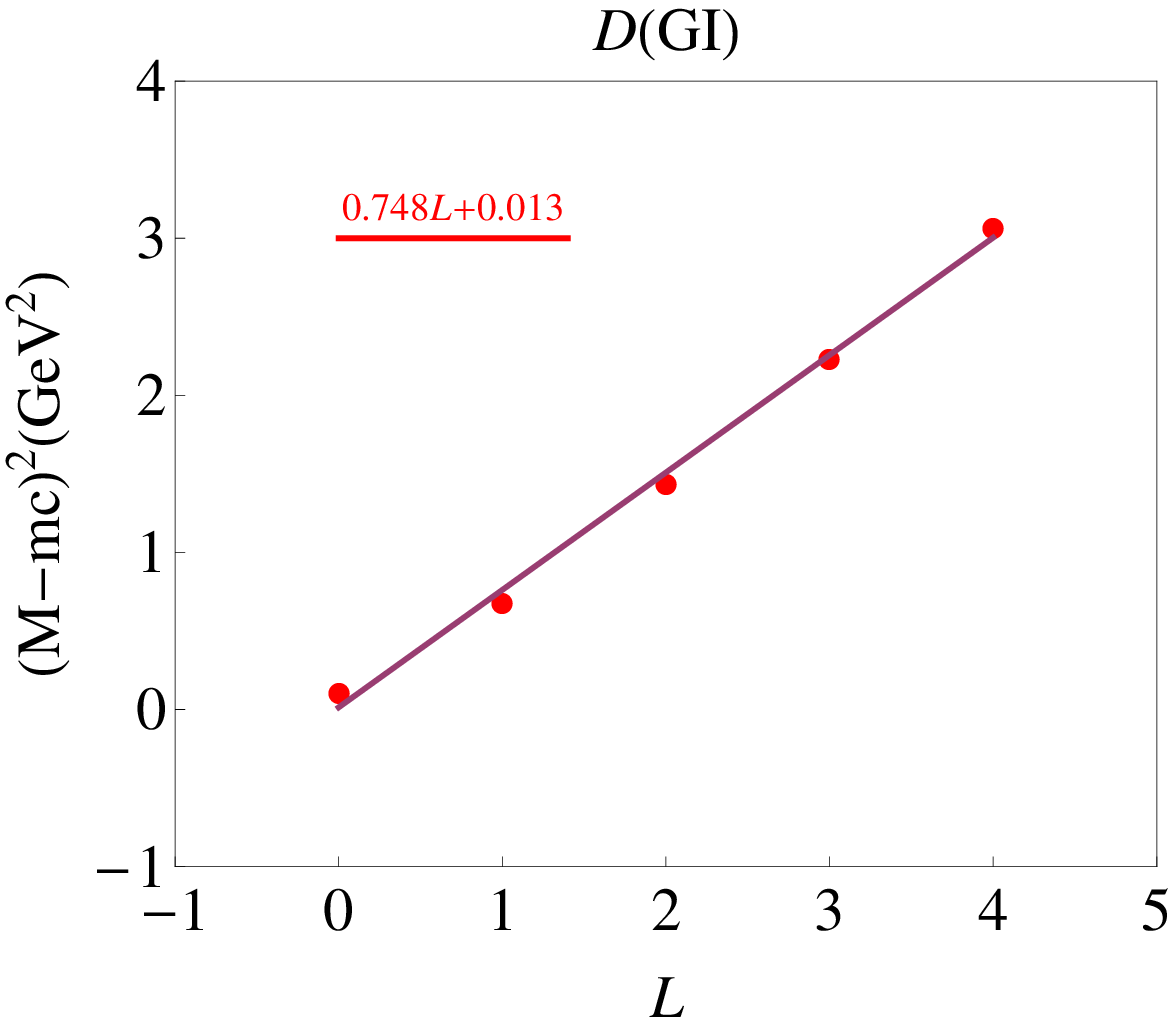} \;\;
		\includegraphics[scale=0.5]{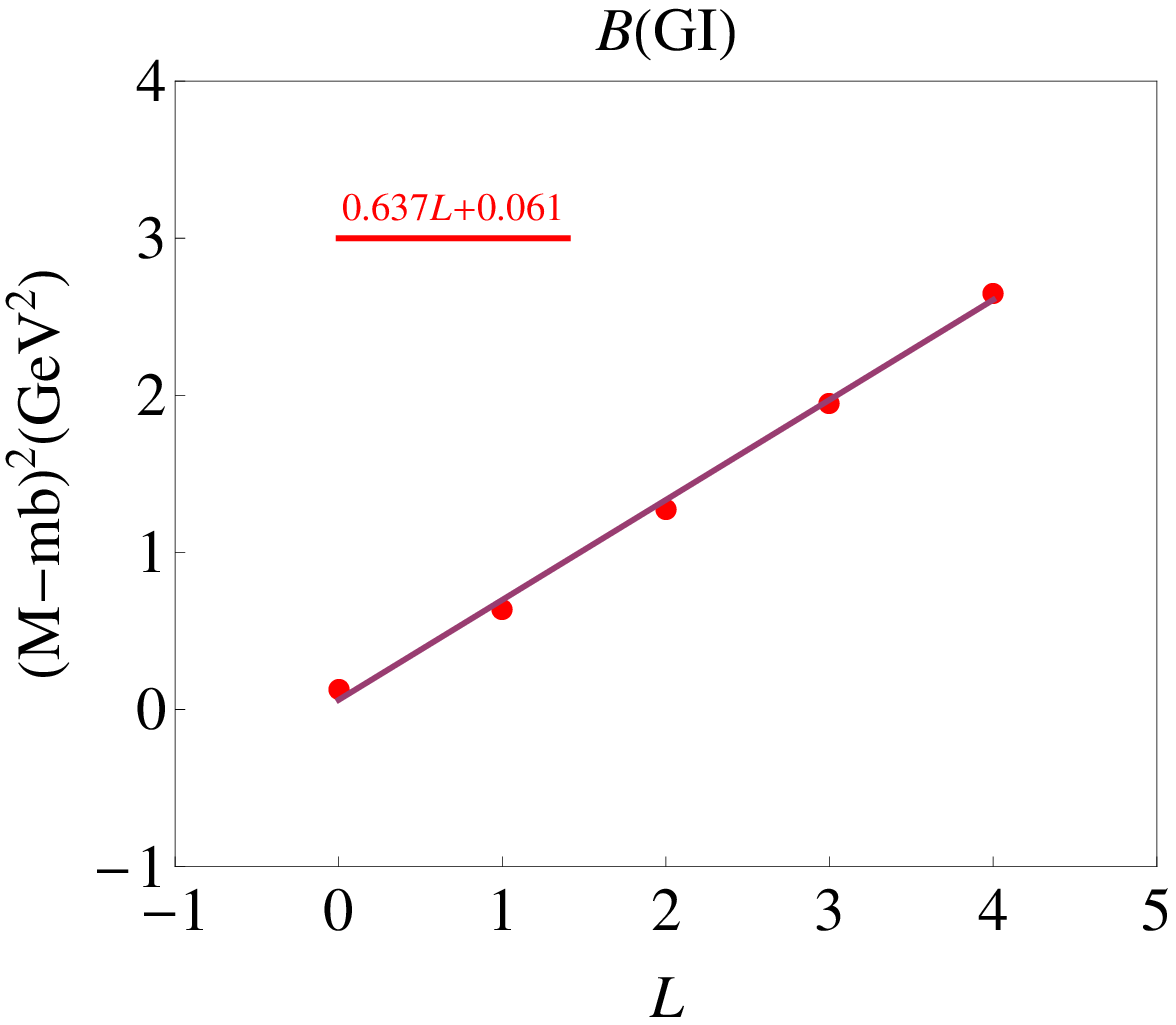}
		\caption{(color online). Plots of values produced by the Godfrey-Isgur model similar to Fig. \ref{expfig}. The best fit lines are given with equations.}\label{modelfig}
	\end{center}
\end{figure}

According to Eq. (\ref{eq:CF2}), we plot figures for $D$ and $B$ mesons for experimental data \cite{Agashe:2014kda} as well as some model calculations given by Godfrey and Isgur \cite{Godfrey:1985xj}, which are given in Figs. \ref{expfig} and \ref{modelfig}. To compare Eq. (\ref{eq:CF1}) with Eq. (\ref{eq:CF2}), we give numerical value of the coefficient of $L$ in Eq. (\ref{eq:CF1}) obtained by Afonin,
\begin{eqnarray}
  M^2 = 1.103L + \cdots.
\end{eqnarray}
which is taken from Table 4 of Ref. \cite{Afonin:2007jd}.

In summary, looking at Figs. \ref{expfig} and \ref{modelfig} together with linear equations written on figures, we can conclude that the string picture for heavy-light mesons well works and supports an approximate rotational symmetry of heavy-light mesons claimed in the previous paper \cite{Matsuki:2016hzk}.

\section*{Acknowledgements}

This work is partly supported by the National
Natural Science Foundation of China under Grant
No. 11475192, as well as supported, in part, by the
DFG and the NSFC through funds provided to the Sino-German
CRC 110 Symmetries and the Emergence of Structure in QCD.
This work is also supported by China Postdoctoral Science Foundation under Grant No. 2016M601133.
T. Matsuki wishes to thank Yubing Dong for his kind hospitality at IHEP, Beijing,
where a part of this work was carried out.


\begin{thebibliography}{}
%
%

\bibitem{Matsuki:2016hzk}
  T.~Matsuki, Q.~F.~L\"u, Y.~Dong and T.~Morii,
  ``Approximate degeneracy of heavy-light mesons with the same $L$,''
  Phys.\ Lett.\ B {\bf 758}, 274 (2016)
  [arXiv:1602.06545 [hep-ph]].
  
\bibitem{Nambu:1974zg} 
  Y.~Nambu,
  ``Strings, Monopoles and Gauge Fields,''
  Phys.\ Rev.\ D {\bf 10}, 4262 (1974).
  doi:10.1103/PhysRevD.10.4262

\bibitem{Afonin:2007jd} 
  S.~S.~Afonin,
  Mod.\ Phys.\ Lett.\ A {\bf 22}, 1359 (2007)
  doi:10.1142/S0217732307024024
  [hep-ph/0701089].

\bibitem{Afonin:2007aa}
  S.~S.~Afonin,
  ``Properties of new unflavored mesons below 2.4-GeV,''
  Phys.\ Rev.\ C {\bf 76}, 015202 (2007)
  [arXiv:0707.0824 [hep-ph]].

\bibitem{Afonin:2013hla}
  S.~S.~Afonin and I.~V.~Pusenkov,
  ``Note on universal description of heavy and light mesons,''
  Mod.\ Phys.\ Lett.\ A {\bf 29}, 1450193 (2014)
  [arXiv:1308.6540 [hep-ph]].

\bibitem{Agashe:2014kda}
  K.~A.~Olive {\it et al.} [Particle Data Group Collaboration],
  ``Review of Particle Physics,''
  Chin.\ Phys.\ C {\bf 38}, 090001 (2014).


\bibitem{Godfrey:1985xj}
  S.~Godfrey and N.~Isgur,
  ``Mesons in a Relativized Quark Model with Chromodynamics,''
  Phys.\ Rev.\ D {\bf 32}, 189 (1985).

\end{thebibliography}
\end{document}